\documentclass[fleqn,12pt,twoside]{article}
\usepackage{espcrc1}
\usepackage{graphicx}
\usepackage{wrapfig}
\usepackage{floatflt}
\usepackage[figuresright]{rotating}

\newcommand{\pt}{\mbox{$p_T$}}
\newcommand{\dphi}{\mbox{$\Delta\phi$}}
\newcommand{\deta}{\mbox{$\Delta\eta$}}

\title{Charged Particle Angular Correlations from Leading Photons at RHIC}

\author{M. Chiu\address{Department of Physics, Columbia University,
        New York, NY 10027, USA},
        for the PHENIX Collaboration\footnote{For the full PHENIX author
        list and acknowledgements, see Appendix "Collaborations" of this volume.}}
       
\begin{document}

\maketitle

\begin{abstract}
We report on the observation of jet-like azimuthal and pseudorapidity correlations
between the highest \pt\ photon (leading photon) and charged hadrons produced in both p-p
and Au-Au collisions at $\sqrt{s_{NN}} = 200$ GeV.
\end{abstract}

\section{Introduction}

The behavior at high-\pt\ in heavy ion collisions
at RHIC is currently not well understood.  Two interesting observations have been
made so far.  First, there is a significant suppression in the yield of high-\pt\
particles relative to the expected yield from binary scaling \cite{Adcox:2001jp}.
Second, high-\pt\ particles have a strong $v_{2}$ correlation to the reaction plane which
is constant with \pt \cite{Adler:2002ct}.  There have been many proposed
theoretical explanations for these observations, such as parton energy
loss \cite{Levai:2001dc,Vitev:2001zn},
gluon saturation \cite{Kharzeev:2002pc}, or correlations from minijets \cite{Kovchegov:2002nf}, but at this time
there is no consensus on the explanation for the above observations.

To better study the origin of these high-\pt\ particles, we study in
detail the angular correlation of charged hadrons from a leading photon
for those events with photons above 2.5 GeV/c in \pt\.  It has been well
established by UA1 and UA2
at the $Sp\bar{p}S$ that high-\pt\ behavior is dominated by hard parton-parton
scattering in nucleon-nucleon collisions, confirming
the previous discovery of this effect at the CERN ISR \cite{Ellis:uc,Jacob:1984vc}.
We expect leading photons in the \pt\ range of our measurement to come predominantly
from the decay of $\pi^{0}$'s.  At high-\pt\ the 
decay photon lies very nearly along the $\pi^{0}$ direction, and since the $\pi^{0}$ comes
from the fragmentation of a jet, the photon is strongly correlated with the jet axis.
By examining the yield of charged hadrons
as a function of $\Delta\phi$ and $\Delta\eta$ with respect to the leading photon in p-p, we
can measure the correlation from the fragmentation of the jets in the near and back-to-back
jet cones.  Then, using the p-p analysis as a reference 
we can check whether similar correlations can be seen in Au-Au, which
would provide evidence for hard scattering behavior in high energy heavy ion collisions.

\section{Analysis}

For this analysis, we use the two central arms of the PHENIX detector.  Each arm covers 
$90^{\circ}$ in $\phi$ and $\pm0.35$ in $\eta$.  The arms consist
of inner tracking (Drift Chambers, Pad Chambers) and particle ID (RICH)
detectors for measuring charged particle type and momenta, followed by
either Lead-Scintillator or Lead-Glass Electromagnetic Calorimeters (EMCal)
for measuring photon
and electron energies.  In addition there are Beam-Beam Counters (BBC) and Zero-Degree
Calorimeters (ZDC) at forward rapidities for global event characterization
and triggering.

The p-p data presented here were obtained from 59,306 events triggered at Level-1
via either the BBC or the EMCal 2x2 and 4x4 tower sum triggers.
The Level-1 triggers are described elsewhere in these proceedings \cite{Torii}.
Our Au-Au sample consists of 170,910 triggered events from scanning
through the PHENIX minimum-bias trigger (BBC and ZDC) plus another 145,285 events from data
obtained with our Level-2 EMCal Tile Trigger.  We take only
those triggered events which have an offline reconstructed photon at a \pt\ above 2.5 GeV/c.
Photons are identified in the EMCal using a combination of Time-of-Flight, shower shape,
and charge veto cuts.
The momenta of the charged particles are reconstructed using the Drift Chamber.
Track quality is enforced by matching to hits in the outer Pad Chambers and the EMCal.
We also apply a RICH veto to remove background electrons.

We correct for the acceptance of the leading-photon/charged-hadron pair by using
mixed events to estimate the correction function in \dphi\ and \deta.
We take the leading photon from the event and calculate $\Delta\phi$ (or $\Delta\eta$)
to a charged hadron taken from another event.  The mixed event distribution
is proportional to the average pair acceptance Acc(\dphi) (or Acc(\deta)),
leaving only an overall efficiency, $\langle\epsilon\rangle$,
which can be computed via Monte Carlo.

\section{Results}

\subsection{p-p}

\begin{figure}
\begin{minipage}[c]{.45\textwidth}
\caption{Per triggered event differential distribution in $\Delta\phi$ for
associated charged hadrons with \pt\ between 2 and 4 GeV/c relative
to a trigger photon in p-p collisions at $\sqrt{s_{NN}} = 200$ GeV. The flat
background from the fit has been subtracted, leaving only the jet-like excess.}
\label{fig:pp_dphi}
\end{minipage}
\begin{minipage}[c]{.5\textwidth}
\includegraphics*[scale=.38]{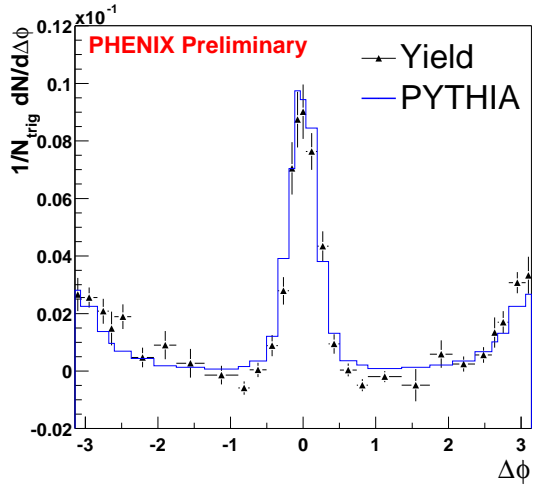}
\end{minipage}
\end{figure}

Fig. \ref{fig:pp_dphi} shows the acceptance-corrected $\Delta\phi$ distribution for
the associated charged particles.  The p-p $\Delta\phi$ distribution at RHIC
includes many of the qualitative features expected from jets, including a large near
angle peak and a smaller peak in the back-to-back region at 180 degrees.
The back-to-back signal is lower in magnitude due to a limited $\eta$ acceptance,
and is broadened from the effects of intrinsic $k_{T}$.

The p-p azimuthal correlation is well described by a fit to Eqn. \ref{ppfit},
which consists of a flat background distribution plus the correlation predicted
by PYTHIA 6.131 (with the default settings) \cite{Sjostrand:1993yb}:
\begin{equation}
\label{ppfit}
\frac{1}{N_{trig}}\frac{dN_{ch}}{d\Delta\phi} =
Acc(\Delta\phi)\langle\epsilon\rangle\left(a_{bkg} +
a_{pythia}\frac{1}{N_{pythia}}\frac{dN_{ch}^{pythia}}{d\Delta\phi}\right)
\end{equation}
In the fit we allow the background level and PYTHIA amplitude to vary.
We checked that within the PYTHIA simulation events with a high-\pt\
photon are dominated by hard scattering and resonance contributions to
$\Delta\phi$ correlations are small.

\subsection{Au-Au}

\begin{figure}
  \begin{minipage}[t]{0.5\linewidth}
  \includegraphics*[scale=.36]{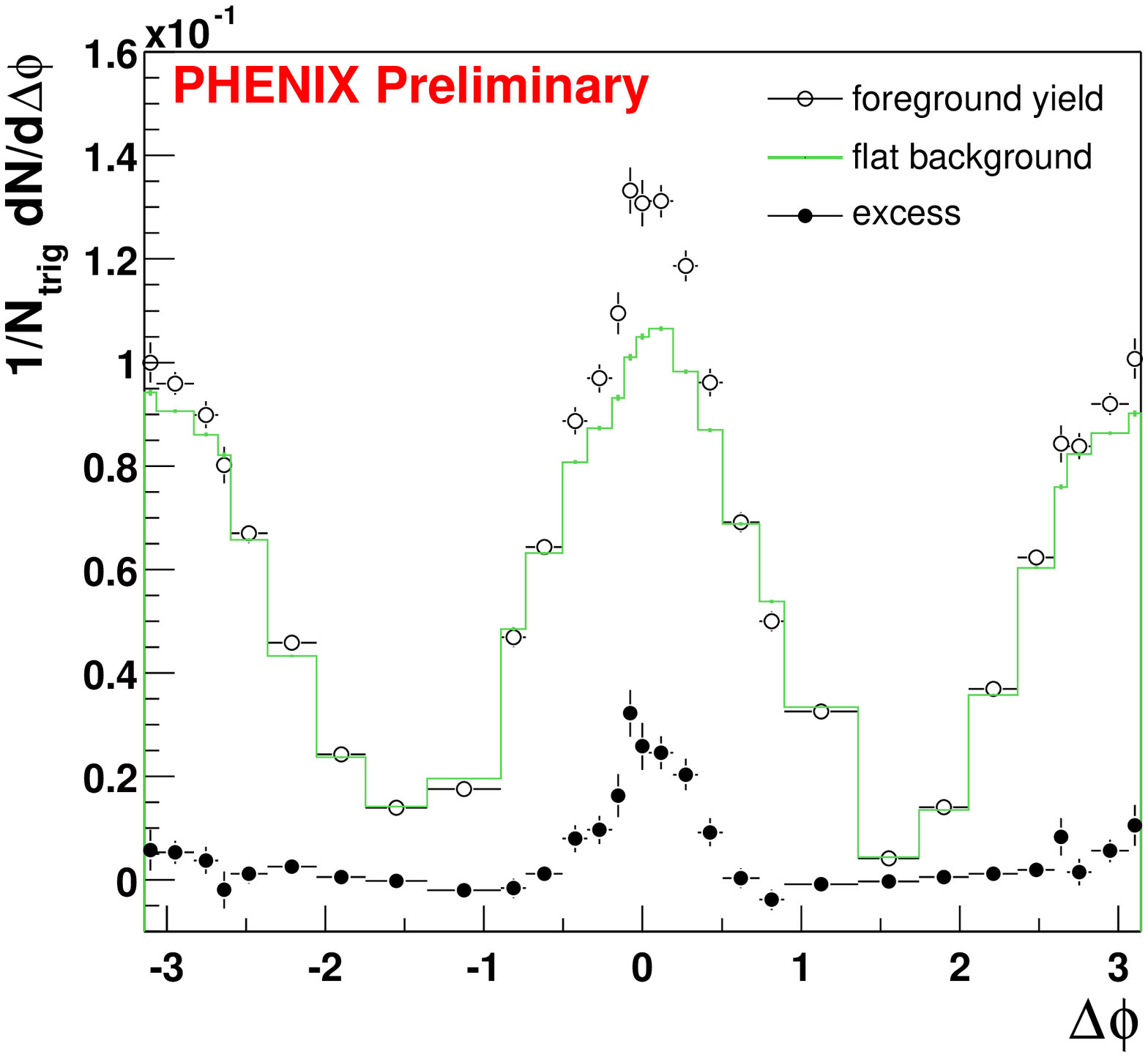}
  \end{minipage}%
  \hspace{.2in}%
  \begin{minipage}[t]{0.5\linewidth}
  \includegraphics*[scale=.36]{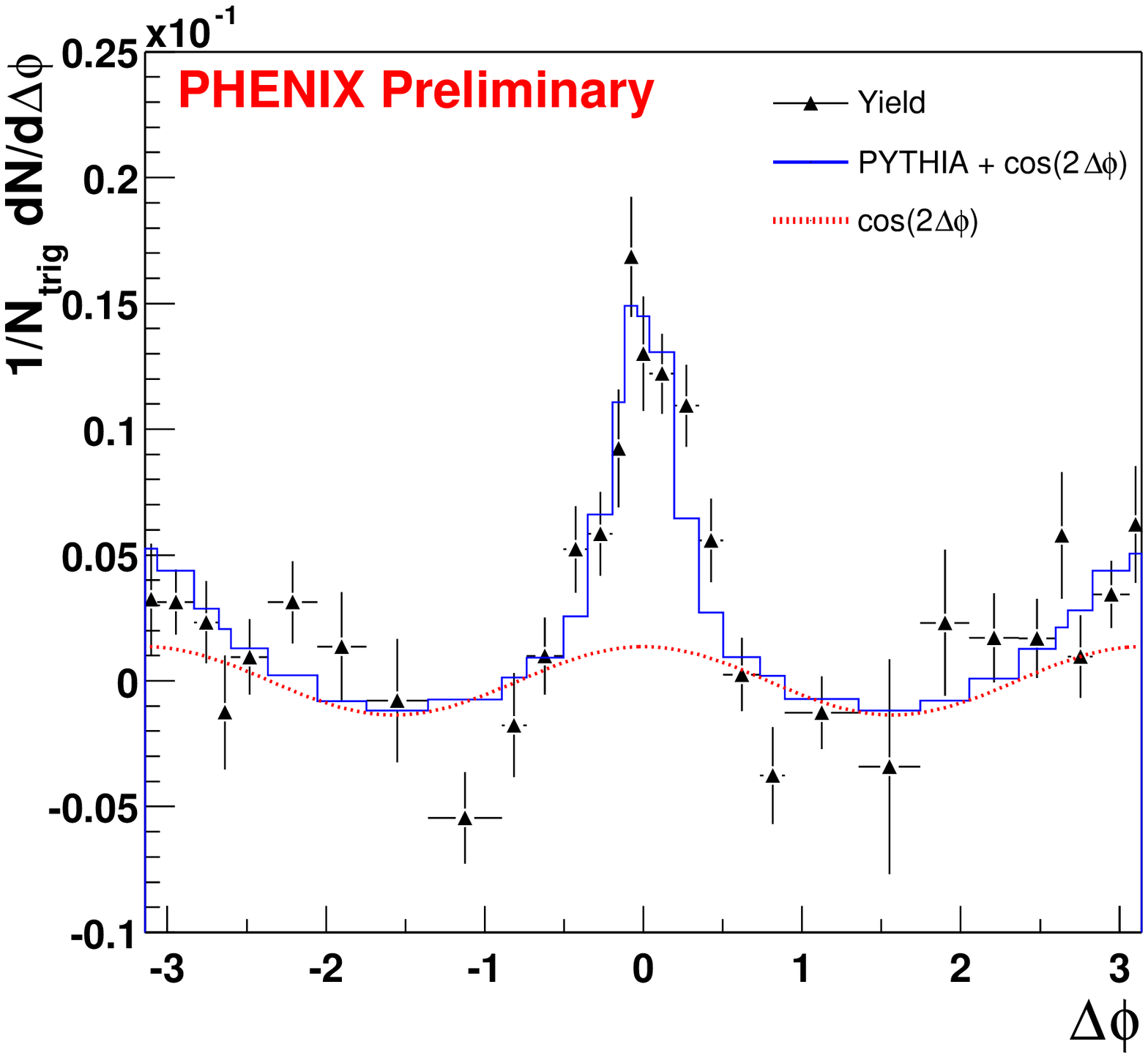}
  \end{minipage}
  \caption{(a) Uncorrected differential yield of associated charged hadrons with
\pt\ between 2 and 4 GeV/c in Au-Au at $\sqrt{s_{NN}} = 200$ GeV, for 20-40\% centrality.
(b) Acceptance-corrected yield and fit to PYTHIA plus an elliptic flow term.  The
flat background has been subtracted to emphasize the fit to the excess.}
  \label{fig:auaudata}
\end{figure}

Perhaps surprisingly, even in the high multiplicity environment
of Au-Au collisions, the azimuthal correlations from jet fragmentation can be
resolved above the background.  In fig.
\ref{fig:auaudata}a, we plot the raw distribution of charged hadrons into the PHENIX Central
Arm acceptance, a flat background folded with the PHENIX acceptance,
and the subtracted distribution.
There is a clear and statistically significant excess above the flat background.
We assume that the contributions to the shape of the excess come from a jet-like
signal plus a $cos(2\Delta\phi)$ term from elliptic flow:
\begin{equation}
\frac{1}{N_{trig}}\frac{dN_{ch}}{d\Delta\phi} =
Acc(\Delta\phi)\langle\epsilon\rangle(a_{bkg} + a_{flow}cos(2\Delta\phi)
+ a_{pythia}\frac{1}{N_{pythia}}\frac{dN_{ch}}{d\Delta\phi})
\label{auaufit}
\end{equation}
If the jet and elliptic flow are not coupled, then $a_{flow} = a_{bkg}*2v_{2}^{trig}v_{2}^{ch}$.

Fig. \ref{fig:auaudata}b shows the result of this fit.  There is good agreement
between the data and our functional fit.  Further checks were made to test whether
the correlations are consistent with those from jets. By studying the
distributions in $\Delta\phi$ for cuts in $|\Delta\eta|>0.35$ and $|\Delta\eta|<0.35$,
we can check whether the particles are simultaneously correlated in \deta\ and \dphi.
For cuts of $|\Delta\eta|>0.35$, which should exclude many particles in a jet cone, we see
a suppression of 5 in the jet-like amplitude relative to that from a cut of
$|\Delta\eta|<0.35$,
consistent with the suppression obtained in the p-p analysis.
We also studied the $\Delta\eta$ distribution directly.  For the near side
($|\Delta\phi|<0.5$), there is a strong peak in \deta\ with a similar width to
the $\Delta\phi$ correlation, while for the away side ($|\Delta\phi|>2.5$)
we see no correlation with \deta.
These studies show that the angular correlations are peaked in a cone at 
\deta\ and \dphi\ angles near to the lead photon, and also peaked around
$\dphi = 180^{\circ}$ but flat in \deta\ on the away side. This behavior
is consistent with results from an identical analysis of our p-p data.

\begin{wrapfigure}[18]{l}{0.50\textwidth}
\includegraphics*[scale=0.35]{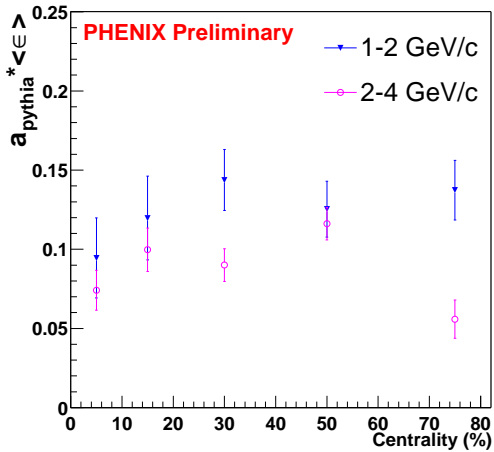}
\caption{Fit values for the jet-like amplitude in Au-Au at
$\sqrt{s_{NN}} = 200$ GeV. Errors are statistical only.}
\label{fig:apythia}
\end{wrapfigure}

In fig. \ref{fig:apythia}, we plot the fit value for the jet-like amplitude
at various centralities and for two \pt\ bins.  The fit values show no strong
centrality dependence, though at this time systematic errors have not been computed
and are under investigation.

\section{Conclusions}

We have studied the azimuthal correlations between leading photons with $\pt>2.5$ GeV/c
and associated charged hadrons in $\sqrt{s}=200$ GeV p-p collisions at RHIC, and
find good agreement with hard scattering expectations as described by a QCD based
Monte Carlo (PYTHIA).  In addition, we find an excess above a constant background in
Au-Au collisions at the same incident energy per nucleon and for different centrality classes,
and find that this excess is described well by a fit to jet-like correlations predicted
by PYTHIA plus an additional flow-like $cos(2\Delta\phi)$ modulation.

\end{document}